\documentclass{rmf-d}
\usepackage{nopageno,rmfbib,multicol,times,epsf,amsmath,amssymb,cite}
\usepackage{supertabular}
\usepackage[latin1]{inputenc}
\usepackage[]{caption2}
\usepackage{graphics}
\usepackage[demo]{graphicx}
\usepackage{hyperref}
\usepackage{lipsum}
\usepackage{wrapfig,blindtext}
\usepackage{comment}
\usepackage{tabularx}
\usepackage{float}
\newenvironment{figurehere}
 {\def\@captype{figure}}
 {}
\makeatother
\def\rmfcornisa{Research OR Education of Physics \hfill\rmf\ 
}

\def\rmfcintilla{{\it Rev.\ Mex.\ Fis.\/}    }
\clearpage \rmfcaptionstyle 
\def\s0{\hspace{-0.20cm} & \hspace{-0.20cm}}
\usepackage{slashed}

\begin{document}
\title{Gluon transversity and TMDs for spin-1 hadrons
\vspace{-6pt}}
\author{S. Kumano$^{a,b}$ and Qin-Tao Song$^{c,d}$}
\address{
$^a$ KEK Theory Center,
             Institute of Particle and Nuclear Studies, 
             KEK,
             Oho 1-1, Tsukuba, Ibaraki, 305-0801, Japan \\
$^b$ J-PARC Branch, KEK Theory Center,
             Institute of Particle and Nuclear Studies, KEK, 
           and Theory Group, Particle and Nuclear Physics Division, 
           J-PARC Center, 
           Shirakata 203-1, Tokai, Ibaraki, 319-1106, Japan \\
$^c$ School of Physics and Microelectronics, Zhengzhou University, 
             Zhengzhou, Henan 450001, China \\
$^d$ CPHT, CNRS, Ecole Polytechnique, Institut Polytechnique de Paris,
     Route de Saclay, 91128 Palaiseau, France
}
\maketitle
\recibido{13 January 2022}{day month 2022\vspace{-12pt}}

\vspace{-0.10cm}
\begin{abstract}
\vspace{1em}   
We explain a gluon transversity,
transverse-momentum-dependent parton distribution functions (TMDs),
and parton distribution functions (PDFs)
for spin-1 hadrons. 
The gluon transversity exists in hadrons with spin more 
than or equal to one, and it does not exist in the spin-1/2 nucleons.
Since there is no direct contribution from the nucleons, 
it is an appropriate quantity to probe an exotic component
in the spin-1 deuteron beyond a simple bound system of the nucleons.
We show how the gluon transversity can be measured at hadron accelerator
facilities by the Drell-Yan process in addition to lepton-accelerator 
experiments.
Next, possible TMDs are explained for the spin-1 hadrons at the twists 3 and 4
in addition to twist-2 ones by considering tensor polarizations.
We found that 30 TMDs exist in the tensor-polarized spin-1 hadron
at the twists 3 and 4 in addition to 10 TMDs at the twist 2.
There are 3 collinear PDFs at the twists 3 and 4.
We also indicate that the corresponding TMD fragmentation functions exist 
at the twists 3 and 4.
Due to the time-reversal invariance in the collinear PDFs,
there are new sum rules on the time-reversal odd TMDs.
In addition, we obtained a useful twist-2 relation, a sum rule,
and relations with multiparton distribution functions
by using the operator product expansion and the equation of
motion for quarks.
These findings are valuable for experimental investigations on
polarized deuteron structure functions in 2020's and 2030's
at world accelerator facilities.
\vspace{1em}
\end{abstract}
\vspace{-0.18cm}

\keys{QCD, quark, gluon, spin-1 hadron, structure function, transversity  \vspace{-4pt}}
\vspace{-0.18cm}
\pacs{\bf{\textit{12.38.-t, 13.60.Hb, 13.88.+e, 24.85.+p}}    
\vspace{-4pt}}
\begin{multicols}{2}

\section{Introduction}
\vspace{-0.10cm}

In spin-1 hadrons, there are structure functions in addition to
the ones of the spin-1/2 nucleons, and they are related to 
their tensor polarizations. These new structure functions were proposed
in 1980's; however, experimental progress was rather slow except
for the HERMES $b_1$ measurement in 2005 \cite{Airapetian:2005cb}.
In spite of this situation, we have a bright future prospect
because there are experimental projects in 2020's and 2030's
to investigate polarized deuteron structure functions 
at various accelerator facilities, such as 
Thomas Jefferson National Accelerator Facility (JLab), 
Fermilab (Fermi National Accelerator Laboratory), 
Nuclotron-based Ion Collider fAcility (NICA),
LHC (Large Hadron Collider)-spin,
and electron-ion colliders (EIC, EicC)
\cite{spin-1-exp}. 
Therefore, time has come to investigate them theoretically
before experimental measurements.

In this report, we discuss a gluon transversity, especially how to 
find it in a proton-deuteron Drell-Yan process. 
There are already significant works on a quark transversity 
both theoretically and experimentally.
However, there is no experimental measurement on the gluon transversity
because it does not exist in the spin-1/2 nucleons. 
The gluon transversity is defined by an amplitude with a gluon-spin flip,
namely the difference of two unit of spin ($\Delta s=2$), so that
the hadron spin needs to be larger than or equal to one.
The most stable spin-1 target is the deuteron, which can be used
for experimental measurements.
Since the proton and neutron cannot contribute directly,
the gluon transversity is an appropriate observable to find
any exotic signature in the deuteron beyond the simple bound
system of the nucleons. If a finite distribution is found experimentally,
it could lead a new field of hadron physics.
On this topic, the purpose of our study is to provide a theoretical
formalism for investigating the gluon transversity at hadron accelerators,
for example, by the Drell-Yan process \cite{ks-trans-g-2020,pd-drell-yan} 
as discussed in Sec.\,3, 
whereas the lepton scattering measurement was already considered at JLab
\cite{spin-1-exp,gluon-trans-2}.

The second topic is on
transverse-momentum-dependent parton distribution functions (TMDs)
and parton distribution functions (PDFs)
of tensor-polarized spin-1 hadrons up to twist 4 \cite{ks-tmd-2021} 
as explained in Secs.\,4 and 5.
Polarized PDFs of the nucleons have been investigated
up to twist 4 \cite{tmds-nucleon}; however, 
they were investigated only at the twist-2 level \cite{bm-2000}
until recently for spin-1 hadrons. 
The purpose of our study is to provide full TMDs, PDFs, and fragmentation
functions up to twist 4 for the spin-1 hadrons \cite{ks-tmd-2021}.
Due to the time-reversal (T) invariance in the collinear PDFs,
there are sum rules for T-odd TMD distributions.
For the collinear PDFs, a useful twist-2 relation and a sum rule were
found \cite{ks-ww-bc-2021} in the similar way to 
the Wandzura-Wilczek relation and the Burkhardt-Cottingham sum rule.
Furthermore, the equation of motion for quarks was used for obtaining
relations among the collinear parton- and multiparton-distribution functions 
for spin-1 hadrons \cite{eq-motion}.
We explain these results in this paper.

\vspace{-0.10cm}
\section{Polarizations of spin-1 hadrons}
\label{polarizations}
\vspace{-0.10cm}

Polarizations of spin-1 hadrons are described by the spin vector $\vec S$ 
and tensor $T_{ij}$ defined by the polarization vector $\vec E$ as

\ \vspace{-0.35cm}
\begin{align}
& \vec S
= \text{Im} \, (\, \vec E^{\, *} \times \vec E \,)
= (S_{T}^x,\, S_{T}^y,\, S_L) ,
\nonumber \\[-0.15cm]
& T_{ij} 
 = \frac{1}{3} \delta_{ij} 
       - \text{Re} \, (\, E_i^{\, *} E_j \,) 
\nonumber \\[-0.02cm]
& \ \hspace{-0.15cm}
= \frac{1}{2} 
\left(
    \begin{array}{ccc}
     - \frac{2}{3} S_{LL} + S_{TT}^{xx}    & S_{TT}^{xy}  & S_{LT}^x  \\[+0.20cm]
     S_{TT}^{xy}  & - \frac{2}{3} S_{LL} - S_{TT}^{xx}    & S_{LT}^y  \\[+0.20cm]
     S_{LT}^x     &  S_{LT}^y              & \frac{4}{3} S_{LL}
    \end{array}
\right) ,
\label{eqn:spin-1-vector-tensor}
\end{align}
where $S_{T}^x$, $S_{T}^y$, $S_L$, 
$S_{LL}$, $S_{TT}^{xx}$, $S_{TT}^{xy}$, $S_{LT}^x$, and $S_{LT}^y$
are parameters to express the vector and tensor polarizations.
The polarizations of the spin-1 hadrons, for example the deuteron,
are listed in Table \ref{table:polarizations} by showing 
the polarization $\vec E$ and the polarization parameters
for the longitudinal, transverse, and linear polarizations
of a spin-1 hadron.
The longitidutinal polarizations contain both $S_L$ and $S_{LL}$,
and the transverse ones do $S^i_T$, $S_{LL}$, and $S^{xx}_{TT}$.
It is interesting to see that these polarizations partially have
the tensor polarization parameter $S_{LL}$ and 
that the linear polarization parameter $S^{xx}_{TT}$ is contained
in the transverse polarization.
The linear polarizations are defined by the polarization vector
$\vec E_x  = \left ( \, 1,\, 0,\, 0 \, \right )$ and
$\vec E_y  = \left ( \, 0,\, 1,\, 0 \, \right )$.
They also have the parameter $S_{LL}$ in addition to $S^{xx}_{TT}$
as shown in Table \ref{table:polarizations}, so that the $S_{LL}$ 
terms should be cancelled in order to extract 
the gluon transversity defined in association 
with $S^{xx}_{TT}$.

\vspace{0.20cm}
\footnotesize
\begin{center}
\renewcommand{\arraystretch}{1.6} 
\bottomcaption{Longitudinal, transverse, and linear polarizations
of a spin-1 hadron, polarization vectors, and parameters 
of the spin vector and tensor \cite{ks-trans-g-2020,spin-1-exp}.}
\label{table:polarizations}
\begin{supertabular}{|l|c|ccccc|} \hline
Polarizations & $\vec E$ &  $S_T^x$
     \s0 $S_T^y$ \s0 $S_L$  \s0 $S_{LL}$ \s0 $S_{TT}^{xx}$ \\ \hline
Longitudinal $+z$  & $\frac{1}{\sqrt{2}} (-1,\, -i,\, 0)$  &
       0    \s0   0     \s0  $+$1 \s0 $+\frac{1}{2}$ \s0  0  \\ \hline
Longitudinal $-z$ & $\frac{1}{\sqrt{2}} (+1,\, -i,\, 0)$ &
       0    \s0   0     \s0  $-$1 \s0 $+\frac{1}{2}$ \s0   0  \\ \hline
Transverse $+x$ & $\frac{1}{\sqrt{2}} (0,\, -1,\, -i)$ &
         $+$1    \s0   0     \s0  0  \s0 $-\frac{1}{4}$  \s0  $+\frac{1}{2}$ \\ \hline
Transverse $-x$ & $\frac{1}{\sqrt{2}} (0,\, +1,\, -i)$ &
         $-1$    \s0   0     \s0  0  \s0 $-\frac{1}{4}$ \s0 $+\frac{1}{2}$ \\ \hline
Transverse $+y$ & $\frac{1}{\sqrt{2}} (-i,\, 0,\, -1)$ &
         0   \s0   $+$1      \s0  0  \s0 $-\frac{1}{4}$ \s0 $-\frac{1}{2}$ \\ \hline
Transverse $-y$ & $\frac{1}{\sqrt{2}} (-i,\, 0,\, +1)$ &
         0   \s0   $-1$     \s0  0  \s0 $-\frac{1}{4}$ \s0 $-\frac{1}{2}$ \\ \hline
Linear  $x$  &  $(1,\, 0,\, 0)$ &
         0   \s0   0      \s0  0  \s0 $+\frac{1}{2}$ \s0 $-1$ \\ \hline
Linear  $y$  &  $(0,\, 1,\, 0)$ &
         0   \s0   0     \s0  0  \s0 $+\frac{1}{2}$ \s0 $+1$ \\ \hline
\end{supertabular}
\end{center}
\normalsize

\section{Gluon transversity in Drell-Yan process}
\label{gluon-transversity}

The gluon transversity has not been measured yet, although
there are global analysis results on the quark transversity.
In principle, it exists in the spin-1 deuteron although
it does not for the spin-1/2 nucleons, so that it is 
a unique quantity to probe a new hadronic physics
within the deuteron.

The gluon transversity $\Delta_T g$ is defined by the matrix element
between the linearly polarized ($E_x$) deuteron as 
\cite{ks-trans-g-2020} 
\begin{align}
\Delta_T g (x) 
& = \varepsilon_{TT,\alpha\beta}
\int \frac{d \xi^-}{2\pi} \, x p^+ \, e^{i x p^+ \xi^-}
\nonumber \\[-0.10cm]
& \ \hspace{0.50cm}
\times 
\langle \, p \, E_{x} \left | \, A^{\alpha} (0) \, A^{\beta} (\xi)  
\right | p \, E_{x} \, \rangle 
_{\xi^+=\vec\xi_\perp=0}  ,
\label{eqn:pdf-definitions}
\end{align}
where $x$ is the momentum fraction for a gluon, 
$\varepsilon_{TT}^{\alpha\beta}$ is the transverse parameter 
given by $\varepsilon_{TT}^{11}=+1$ and $\varepsilon_{TT}^{22}=-1$,
$\xi$ is the space-time coordinate expressed by the lightcone 
coordinates $\xi^\pm = (\xi^0 \pm \xi^3)/\sqrt{2}$ and $\vec\xi_\perp$,
$p$ is the deuteron momentum, and $A^\mu$ is the gluon field.
It is expressed by the gluon distribution difference as
\begin{align}
\Delta_T g (x) = g_{\hat x/\hat x} (x) - g_{\hat y/\hat x} (x) ,
\label{eqn:gluon-transversity-linear}
\nonumber \\[-0.70cm]
\end{align}
where $\hat y/\hat x$ is the gluon linear polarization
$\varepsilon_y$ in the deuteron with the polarization $E_x$.
In terms of parton-hadron forward scattering amplitudes
$A_{\Lambda_i \lambda_i ,\, \Lambda_f \lambda_f}$
with the initial and final hadron helicities
$\Lambda_i$ and $\Lambda_f$ and parton ones
$\lambda_i$ and $\lambda_f$,
the gluon transversity is given by
\begin{align}
\Delta_T g (x)  \sim \text{Im} \, A_{++,\, - \hspace{0.03cm} -} \ .
\label{eqn:delta-deltaT-amplitudes}
\end{align}
Namely, it is defined by the amplitude with the gluon helicity flip,
so that the change of two spin units ($\Delta s=2$) is needed
between the initial and final states. It is the reason why
the spin-1/2 nucleons cannot accommodate this distribution.

The gluon transversity will be measured at charged-lepton scattering
by looking by the angle dependence of the deuteron spin
in the cross section \cite{spin-1-exp}. It is the angle between
the lepton-scattering plan and the target-spin orientation.
The intension of our studies is to make the measurement possible
at hadron accelerator facilities by supplying a theoretical formalism
for the Drell-Yan process \cite{ks-trans-g-2020,pd-drell-yan}.
As an example, the proton-deuteron Drell-Yan process was investigated
because it is possible at Fermilab. The formalism details are explained 
in the paper \cite{ks-trans-g-2020}, where the cross section of
$p(A)+d(B) \to \mu^+ \mu^- +X$ is given by
the difference $d\sigma (E_x) - d\sigma (E_y)$ as 
\begin{align}
& \frac{ d \sigma_{pd \to \mu^+ \mu^- X} }{d\tau \, d \vec q_T^{\, 2} \, d\phi \, dy}
(E_x-E_y )
 = - \frac{\alpha^2 \, \alpha_s \, C_F \, q_T^2}{6\pi s^3} \cos (2\phi) 
\nonumber \\
&
 \times 
\int_{\text{min}(x_a)}^1 \! dx_a 
 \frac{ \sum_{q}  e_q^2 \, x_a \!
 \left[ \, q_A (x_a) + \bar q_A (x_a) \, \right ] x_b \Delta_T g_B (x_b)} 
 { (x_a x_b)^2 \, (x_a -x_1) (\tau -x_a x_2 )^2} ,
\label{eqn:cross-5}
\end{align}
\ \vspace{-0.45cm} \  

\noindent
by considering the deuteron linear polarizations ($E_x$, $E_y$).
Here, $\tau$ is defined by the dimuon mass or momentum squared as
$\tau=M_{\mu\mu}^2/s=Q^2/s$ with the center-of-mass energy squared $s$,
$\vec q_T^{\,2}$ is the dimuon transverse momentum squared,
 $\phi$ is its azimuthal angle, $y$ is the rapidity 
in the center-of-mass frame,
$\alpha$ is the fine structure constant,
$\alpha_s$ is the QCD running coupling constant,
$C_F$ is the color factor $C_F=(N_c^2-1)/(2N_c)$ with $N_c=3$,
and $e_q$ is the quark charge.
The momentum fraction $x_b$ is given by
$x_b=(x_a x_2 -\tau)/(x_a-x_1)$, and the minimum 
of $x_a$ is $\text{min}(x_a)=(x_1-\tau)/(1-x_2)$
with $x_1 = e^y \sqrt{(Q^2+\vec q_T^{\,2})/s}$
and $x_2 = e^{-y} \sqrt{(Q^2+\vec q_T^{\,2})/s}$.
The $q_A (x_a)$ and $\bar q_A (x_a)$ 
are quark and antiquark distribution functions in the proton,
and $\Delta_T g (x_b)$ is the gluon transversity in the deuteron.

In estimating the cross section numerically, we used
the CTEQ14 PDFs for the unpolarized PDFs of the proton and also
the deuteron by ignoring nuclear corrections.
Since there is no available gluon transversity at this stage,
we assumed it is equal to the longitudinally-polarized gluon
distribution given by the NNPDF1.1; however, it is likely
an overestimation of the cross section.
In Fig.\,1, the polarization asymmetry 
$A_{E_{xy}}\equiv d\sigma (E_x-E_y)/d\sigma (E_x+E_y)$
is shown for the Fermilab kinematics with $p_p = 120$ GeV
by taking $\phi=0$, $y=0.5$, and $q_T=0.5$ or 1.0 GeV
as the function $M_{\mu\mu}^2$.
The asymmetry is typically a few percent. However,
if a finite gluon transversity is found in an experiment,
it could lead to an interesting new hadron physics.
Fortunately, this experiment will be proposed at Fermilab
within the E-1039 collaboration \cite{spin-1-exp}.

\begin{figurehere}
\begin{center}
   \includegraphics[width=6.0cm]{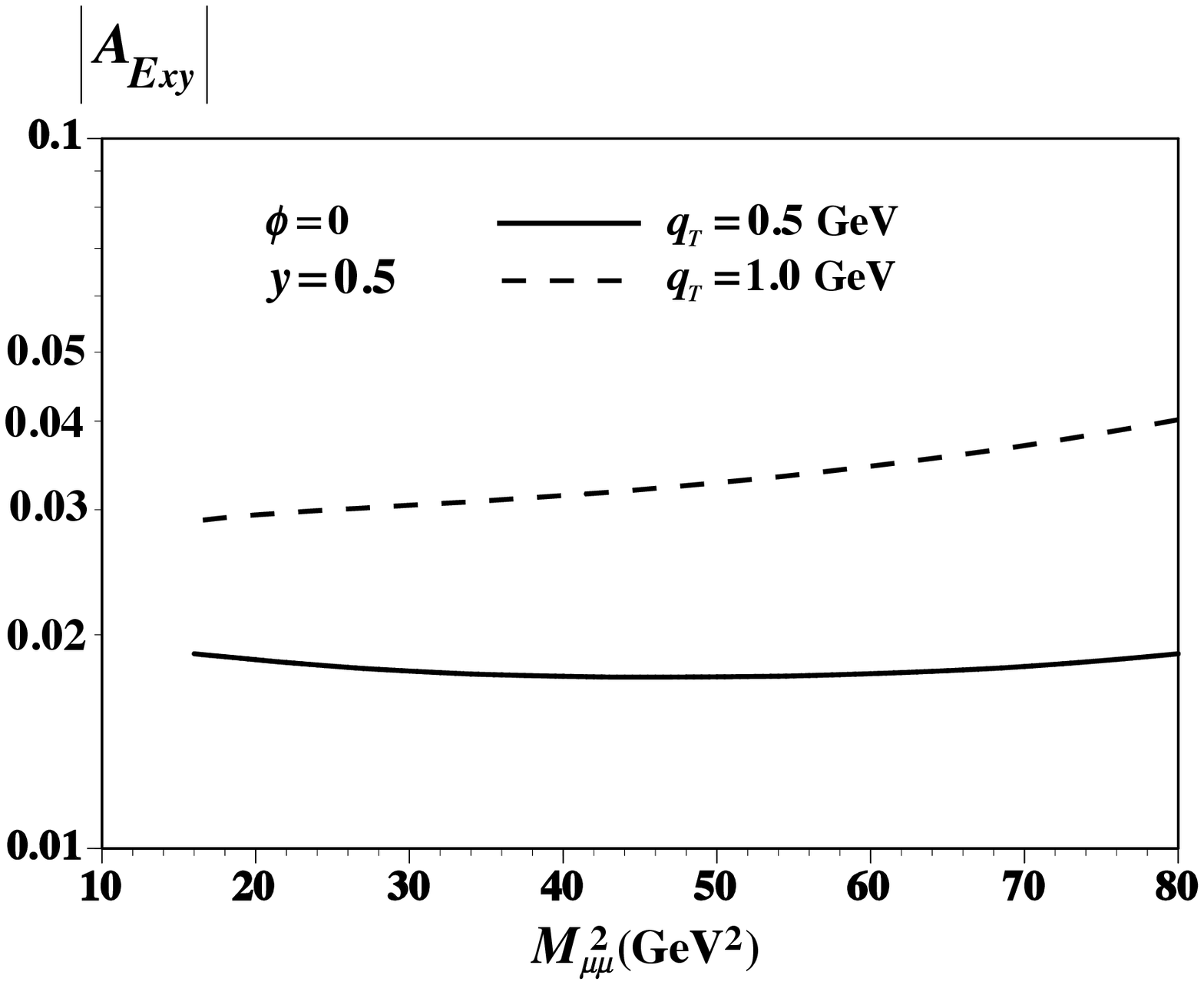} \\
Figure 1. Polarization asymmetry $ | A_{E_{xy}} |$.
\end{center}
\label{fig:asym-ex-ey}
\end{figurehere}

\section{TMDs and PDFs for spin-1 hadrons}
\label{TMDs-up-to-twist-4}

Next, we discuss the TMDs and PDFs at the twists 3 and 4
for tensor-polarized spin-1 hadrons.
Recently, fully consistent investigations have been done for finding
possible twist 3 and twist 4 TMDs and PDFs, whereas the higher-twist
PDFs were found many years ago for the nucleons.
In general, the TMDs and PDFs are defined from the correlation function
$\Phi_{ij}^{[c]}$,
which is the amplitude to extract a parton from a hadron and 
then to insert it into the hadron at a different space-time point:
\begin{align}
& \Phi_{ij}^{[c]} (k, P, T  \, | \, n )
= \int  \! \frac{d^4 \xi}{(2\pi)^4} \, e^{ i k \cdot \xi}
\nonumber \\[-0.15cm]
& \ \hspace{1.7cm}
\times
\langle \, P , T \left | \, 
\bar\psi _j (0) \,  W^{[c]} (0, \xi)  
 \psi _i (\xi)  \, \right | P, \,  T \, \rangle .
\label{eqn:correlation-q}
\end{align} 
Here, $k$ and $P$ are quark and hadron momenta, $T$ indicates
the tensor polarization of a spin-1 hadron, 
$n$ is the lightcone vector $n^\mu =(1,0,0,-1)/\sqrt{2}$,
$\xi$ is a space-time coordinate, 
$\psi$ is the quark field, 
and $W^{[c]} (0, \xi)$ is the gauge link with the integral path $c$.

This correlation function is expanded in a Lorentz invariant way
with the constraints of the Hermiticity and parity invariance.
The time-reversal invariance does not have to be satisfied in the TMD 
level due to the existence of the color flow given by the gauge link;
however, it is imposed in the collinear PDFs. Then, we obtain
\cite{ks-tmd-2021}
\begin{align}
\Phi(k, P, T \, | n) & = \frac{A_{13}}{M}  T_{kk} 
+ \cdots
+ \frac{A_{20}}{M^2} \varepsilon^{\mu\nu P k}  \gamma_{\mu} \gamma_5 T_{\nu k}
\nonumber \\
&
+ \frac{B_{21}M}{P\cdot n} T_{kn}  
+ \cdots
+ \frac{B_{52}M}{P\cdot n } \sigma_{\mu k}  T^{\mu n} ,
\label{eqn:cork4}
\end{align} 
for the tensor polarization part.
The twist-2 expression was given in Ref.\,\cite{bm-2000} for spin-1 hadrons.
Higher-twist expressions were investigated for the spin-1/2 nucleons 
in Ref.\,\cite{tmds-nucleon} by including the lightcone vector $n$
to accommodate twist-3 and 4 effects.
In the same way, we included $n$ terms for the spin-1 hadrons
for defining the higher-twist TMDs and PDFs.
Here, $A_i$ and $B_i$ are expansion coefficients,
the tensor polarization is expressed by $T^{\mu\nu}$,
and the contraction $X_{\mu k} \equiv X_{\mu \nu} k^{\nu}$ is used.
The TMDs are given by integrating the function over the quark momenta as
\begin{align}
\Phi^{[c]} (x, k_T, P, T ) & = \! \int \! dk^+ dk^- \, 
               \Phi^{[c]} (k, P, T  \, |n ) \, \delta (k^+ \! -x P^+) .
\label{eqn:correlation-tmd}
\\[-1.00cm] \nonumber
\end{align}

The TMDs and collinear PDFs are defined by traces of 
the correlation functions with $\gamma$ matrices ($\Gamma$) as
$ \Phi^{\left[ \Gamma \right]} \equiv 
\text{Tr} \left[ \, \Phi \Gamma \, \right] /2 $.
The twist-2 TMDs were defined
by the traces $\Phi^{ [ \gamma^+ ] }$,
$\Phi^{ [ \gamma^+ \gamma_5 ] }$, and
$\Phi^{ [ i \sigma^{i+} \gamma_5 ] }$
(or $\Phi^{ [ \sigma^{i+} ] }$) \cite{bm-2000}.
The twist-3 TMDs were obtained by 
$\Phi^{ [ \gamma^i ] }$,
$\Phi^{\left[{\bf 1}\right]}$,
$\Phi^{\left[i\gamma_5\right]}$
$\Phi^{ [\gamma^{i}\gamma_5 ]}$
$\Phi^{ [ \sigma^{ij} ]}$,
and $\Phi^{ [ \sigma^{-+} ] }$,
and the twist-4 TMDs were obtained by
$\Phi^{[\gamma^-]}$,
$\Phi^{[\gamma^- \gamma_5]}$, and $\Phi^{[\sigma^{i-}]}$
\cite{ks-tmd-2021}.
For example, we have
\begin{align}
& 
\Phi^{ [ \gamma^i ] } (x, k_T, T)
= 
\frac{M}{P^+} \bigg [  f^{\perp}_{LL}(x, k_T^{\, 2})  S_{LL} \frac{k_T^i}{M}
\! + \! f^{\,\prime} _{LT} (x, k_T^{\, 2})S_{LT}^i 
\nonumber \\[-0.10cm]
& \ \hspace{1.0cm}
- f_{LT}^{\perp}(x, k_T^{\, 2}) \frac{ k_{T}^i  S_{LT}\cdot k_{T}}{M^2} 
- f_{TT}^{\,\prime} (x, k_T^{\, 2}) \frac{S_{TT}^{ i j} k_{T \, j} }{M} 
\nonumber \\[-0.10cm]
& \ \hspace{1.0cm}
+ f_{TT}^{\perp}(x, k_T^{\, 2}) \frac{k_T\cdot S_{TT}\cdot k_T}{M^2} 
       \frac{k_T^i}{M} \bigg ] ,
\label{eqn:cork-3-1a}
\end{align} 
as the trace for defining some of the twist-3 TMDs.
Instead of the TMDs with $^\prime$, we define other TMDs by
$
F (x, k_T^{\, 2}) \equiv F^{\,\prime} (x, k_T^{\, 2})
 - (k_T^{\, 2} /(2M^2)) \, F^{\perp} (x, k^{\, 2}_T) 
$
where $k_T^{\, 2}= - \vec k_T^{\, 2}$,
so that the TMDs with $^\prime$ may not be used
in actual TMD lists.
From these traces, we find that the following tensor-polarized TMDs exist
\cite{ks-tmd-2021}:
\begin{align}
& \text{Twist-2 TMD:}\ \ f_{1LL},\ f_{1LT},\ f_{1TT},\ g_{1LT},\ g_{1TT},
\nonumber \\[-0.20cm]
& \ \hspace{2.3cm}
      h_{1LL}^\perp,\ h_{1LT},\ h_{1LT}^\perp,\ h_{1TT},\ h_{1TT}^\perp ,
\nonumber \\
& \text{Twist-3 TMD:}\ \ f_{LL}^\perp,\ e_{LL},\  
      f_{LT},\ f_{LT}^\perp,\ e_{1T},\ e_{1T}^\perp,\ 
      f_{TT},\ f_{TT}^\perp,
\nonumber \\[-0.20cm]
& \ \hspace{2.3cm}      
      e_{TT},\ e_{TT}^\perp,\ 
      g_{LL}^\perp,\ g_{LT},\ g_{LT}^\perp,\ g_{TT},\ g_{TT}^\perp,
\nonumber \\[-0.20cm]
& \ \hspace{2.3cm}
      h_{1L},\ h_{LT},\ h_{LT}^\perp,\ h_{TT},\ h_{TT}^\perp,
\nonumber \\
& \text{Twist-4 TMD:}\ \ f_{3LL},\ f_{3LT},\ f_{3TT},\ g_{3LT},\ g_{3TT},
\nonumber \\[-0.20cm]
& \ \hspace{2.3cm}
      h_{3LL}^\perp,\ h_{3LT},\ h_{3LT}^\perp,\ h_{3TT},\ h_{3TT}^\perp .
\label{eqn:spin-1-tmds-2-3-4}
\end{align} 
Namely, there are 10, 20, and 10 tensor-polarized TMDs 
at twists 2, 3, and 4, respectively.
These are classified by chiral even/odd and time-reversal even/odd.
Since the time-reversal invariance should be satisfied 
in collinear PDFs by the integral over the transverse momentum $\vec k_T$,
there are sum rules for the T-odd TMDs as
\begin{align}
& \! \int \! d^2 k_T \, h_{1LT} (x, k_T^{\, 2}) 
= \! \int \! d^2 k_T \, g_{LT} (x, k_T^{\, 2}) 
\nonumber \\[-0.20cm]
& \ \ \ 
= \! \int \! d^2 k_T \, h_{LL} (x, k_T^{\, 2}) 
= \! \int \! d^2 k_T \, h_{3LT}(x, k_T^{\, 2})  = 0 .
\label{eqn:TMD-sum}
\end{align} 

The TMD fragmentation functions are also found up to twist 4 \cite{ks-tmd-2021}
simply by changing kinematical variables and function names as
\cite{bm-2000}
\vspace{-0.20cm}
\begin{align}
& \ \hspace{-0.20cm}
\text{Kinematical variables:}   \ \  
x, k_T, S, T, M, n, \gamma^+, \sigma^{i+}
\nonumber \\[-0.15cm]
& \ \hspace{0.5cm}
\Rightarrow \ 
 z, k_T, S_h, T_h, M_h, \bar n, \gamma^-, \sigma^{i-},
\nonumber \\
& \ \hspace{-0.20cm}
\text{Distribution functions:}  \ \ f, g, h, e \hspace{2.30cm}
\nonumber \\[-0.15cm]
& \ \hspace{0.5cm}
\Rightarrow \ 
\text{Fragmentation functions:} \ 
D, G, H, E .
\label{eqn:tmd-fragmentation}
\end{align} 
In addition, if the TMDs are integrated over $\vec k_T$, we obtain
the tensor-polarized PDFs up to twist 4 as
\vspace{-0.20cm}
\begin{align}
& \text{Twist-2 PDF:}\ f_{1LL}, \ 
\text{Twist-3:}\ e_{LL},\ f_{LT}, 
\nonumber \\[-0.15cm]
& \text{Twist-4:}\ f_{3LL}. 
\label{eqn:pdfs-2-3-4}
\nonumber \\[-0.90cm] 
\end{align} 
The collinear fragmentation functions were investigated
in Ref.\,\cite{ji-ffs}.

\section{Useful relations among PDFs and multiparton distribution functions}
\label{useful-relations}

For the new twist-3 PDFs, we derived useful relations.
First, we obtained a twist-2 relation and a sum rule \cite{ks-ww-bc-2021},
analogous to the Wandzura-Wilczek (WW) relation 
and the Burkhardt-Cottingham (BC) sum rule,
for the twist-2 and twist-3
tensor-polarized parton distribution functions $f_{1LL}$ 
and $f_{LT}$, respectively. 
Using the formalism of the operator product expansion and defining
multiparton distribution functions for twist-3 terms,
we obtained the relation
\begin{align}
\! 
f_{LT}(x)= \frac{3}{2} \int^{\epsilon (x)}_x dy \frac{f_{1LL}(y)}{y}
          +\int^{\epsilon (x)}_x dy \frac{f_{LT}^{(HT)}(y)}{y}.
\label{eqn:flt}
\end{align}
Here, $\epsilon (x)$ is defined by $\epsilon (x)=1$ ($-1$) at $x>0$ ($x<0$),
and the last term is the twist-3 effect given by
the multiparton distribution functions.
We define $+$ PDFs by $f^+ (x) = f(x) + \bar f(x)$ 
in the range $0 \le x \le 1$. 
The $f_{1LL}^+$ is the same as $b_1$ with the relation
$b_1^{q+\bar q} = - (3/2) f_{1LL}^+$.
Then, neglecting the higher-twist term,
we obtain
\begin{align}
f_{LT}^+(x)= \frac{3}{2} \int^1_x 
\frac{dy}{y} \, f_{1LL}^+(y) .
\end{align}
Namely, the twist-2 part of $f_{LT}$ is expressed by an integral of
$f_{1LL}$ (or $b_1$).
If the function $f_{2LL}$ is define by
$f_{2LL} = \frac{2}{3} f_{LT}-f_{1LL}$,
it leads to the twist-2 relation similar to the WW relation as
\begin{align}
f_{2LT}^+ (x)=-f_{1LL}^+ (x)+ \int^1_x \frac{dy}{y} f_{1LL}^+ (y) .
\nonumber 
\end{align}
Integrating this equation over $x$, we obtain the BC-like sum rule as
\begin{align}
& \int_0^1 dx \, f_{2LT}^+(x) =0 .
\nonumber 
\end{align}
If the parton-model sum rule for $f_{1LL}$ ($b_1$), 
$\int dx f_{1LL}^+ (x) = 0$ 
($\int dx b_1^{q+\bar q} (x) = 0$) \cite{b1-sum},
is applied by assuming vanishing tensor-polarized antiquark distributions, 
another sum rule exists for $f_{LT}$ itself,
$\int_0^1 dx \, f_{LT}^+(x) =0.$
In deriving these relation, we showed that the following
tensor-polarized multiparton distribution functions exist:
$F_{LT} (x,y),\ G_{LT} (x,y),\ 
H_{LL}^\perp (x,y),\  H_{TT} (x,y) .$

Next, from the equation of motion for quarks, useful relations
were also obtained 
(1) 
for the twist-3 PDF $f_{LT}$,
the trasverse-momentum moment PDF $f_{1LT}^{\,(1)}$, and 
the multiparton distribution functions $F_{G,LT}$ and $G_{G,LT}$;
(2)
for the twist-3 PDF $e_{LL}$, the twist-2 PDF $f_{1LL}$,
and the multiparton distribution function $H_{G,LL}^\perp$ as
\cite{eq-motion} 
\begin{align}
& \! \!
x f_{LT}(x) - f_{1LT}^{\,(1)}(x)
- {\cal P} \! \! \int_{-1}^1 \! \! \! dy \,
\frac{F_{G,LT}(x, y) + G_{G,LT} (x, y) }{x-y} = 0 ,
\nonumber \\
& \! \!
x \, e_{LL}(x) - 2 {\cal P} \! \! \int_{-1}^1 \! \! \! dy \, 
 \frac{H_{G,LL}^\perp (x, y)}{x-y} 
-\frac{m}{M} f_{1LL} (x) 
=0 .
\nonumber 
\end{align}
The transverse-momentum moments of the TMDs are defined by
$f^{\, (1)} (x) = \int \! d^2 k_T 
(\vec k_T^{\,2} / (2 M^2)) \, f(x,k_T^2)$,
${\cal P}$ is the principle integral,
and $m$ is the quark mass.
In addition, the Lorentz-invariance relation was obtained as
\cite{eq-motion}
\begin{align}
\! \! 
\frac{d f_{1LT}^{\,(1)}(x) }{dx}
- f_{LT}(x) + \frac{3}{2} f_{1LL}(x)
- 2 {\cal P}   \! \!
 \int_{-1}^1   \! \! \!  dy \, \frac{F_{G,LT} (x, y)}{(x-y)^2} =0 .
\nonumber 
\end{align}
In these derivations, we also obtained relations
in $F_{D/G,LT} (x, y)$, $G_{D/G,LT} (x, y)$,
$H_{D/G,LL}^\perp (x, y)$, and $H_{D/G,TT} (x, y)$.

\section{Summary}

We explained a possible gluon transversity measurement by the proton-deuteron
Drell-Yan process. Then, possible twist-3 and twist-4 TMDs and PDFs were shown
for tensor-polarized spin-1 hadrons. In addition, the corresponding 
TMD fragmentation functions exist at twists 3 and 4.
A useful twist-2 WW-like relation and a BC-like sum rule were derived
by defining multiparton distribution functions at twist 3.
Furthremore, from the equation of motion for quarks,
the twist-3 PDFs are related to other PDFs and multiparton 
distribution functions, and so called the Lorentz-invariance relation
was also obtained. Since there are various experimental projects
to investigate spin-1 hadrons, these studies should be useful.

\section*{Acknowledgments}
S. Kumano was partially supported by 
Japan Society for the Promotion of Science (JSPS) Grants-in-Aid 
for Scientific Research (KAKENHI) Grant Number 19K03830.
Qin-Tao Song was supported by the National Natural Science Foundation 
of China under Grant Number 12005191, the Academic Improvement Project 
of Zhengzhou University, and the China Scholarship Council 
for visiting Ecole Polytechnique.

\end{multicols}
\medline

\begin{multicols}{2}


\end{multicols}
\end{document}